# Trafficking coordinate description of intracellular transport control of signaling networks

Jose M. G. Vilar[1,2,*] and Leonor Saiz[3,*]
[1] Biophysics Unit (CSIC-UPV/EHU) and Department of Biochemistry and Molecular Biology, University of the Basque Country, P.O. Box 644, 48080 Bilbao, Spain
[2] IKERBASQUE, Basque Foundation for Science, 48011 Bilbao, Spain
[3] Department of Biomedical Engineering, University of California, 451 E. Health Sciences Drive, Davis, CA 95616, USA

**Abstract**
**Many cellular networks rely on the regulated transport of their components to transduce extracellular information into precise intracellular signals. The dynamics of these networks is typically described in terms of compartmentalized chemical reactions. There are many important situations, however, in which the properties of the compartments change continuously in a way that cannot naturally be described by chemical reactions. Here, we develop an approach based on transport along a trafficking coordinate to precisely describe these processes and we apply it explicitly to the TGF-β signal transduction network, which plays a fundamental role in many diseases and cellular processes. The results of this newly introduced approach accurately capture for the first time the distinct TGF-β signaling dynamics of cells with and without cancerous backgrounds and provide an avenue to predict the effects of chemical perturbations in a way that closely recapitulates the observed cellular behavior.**

---

[*] Correspondence: j.vilar@ikerbasque.org or lsaiz@ucdavis.edu

## Introduction

The cellular behavior is coordinated by complex networks of interacting molecules that operate at different levels of organization (1-4). At the cell surface, transmembrane receptors sense extracellular cues and transduce them into precise intracellular signals. These receptors are not just passive signal transducers but are able to process the signals before passing them downstream. In organisms like bacteria, processing is typically done by chemical modifications of the receptors (5-7). In cells of higher organisms, like mammals and other eukaryotes, there are additional layers of control. One such layer is receptor trafficking, which has been shown to strongly regulate signal transduction (8-10). This additional layer endows the receptor level with the ability to detect absolute levels of ligands, temporal changes in ligand concentration, and ratios of multiple ligands (11).

Receptor trafficking has been investigated in detail in many signal transduction pathways, such as those of the epidermal growth factor receptor (EGFR), G protein-coupled receptors (GPCR), and the transforming growth factor β (TGF-β) receptors (12-15). Typically, trafficking and signaling are coupled through the induction of receptor internalization upon ligand binding and receptor activation, as for instance in the EGFR and GPCR pathways. After internalization via endocytosis, receptors can activate different signaling pathways, be modified in specific ways, and be targeted for degradation or recycling back to the plasma membrane. In the case of the TGF-β pathway, receptors are constitutively internalized, even in the absence of ligand (16, 17). The trafficking route that the receptors follow, however, depends on whether or not they are in an active signaling complex with the ligand. Different routes trigger different signaling outcomes and affect how receptors are degraded.

The typical way in which trafficking is analyzed is to consider it as chemical reactions and transport between compartments. Explicitly, given a species $i$ in the compartment $j$ with concentration $x_{i,j}$, the traditional approach considers that the dynamics is given by the reaction equation

$$\frac{dx_{i,j}}{dt} = \sum_{k \neq j} T^i_{j \leftarrow k} x_{i,k} - \sum_{k \neq j} T^i_{k \leftarrow j} x_{i,j} + f_{i,j}(x_{1,j},...,x_{N,j}), \tag{1}$$

where $T^i_{j \leftarrow k}$ is the trafficking rate of species $i$ from compartment $k$ to $j$ and $f_{i,j}$ is the function that gives the change (production or consumption) of $x_{i,j}$ as a result of the reaction between the different species within the compartment. In the case of the TGF-β pathway, the molecular species are receptors and ligand-receptor complexes, which can be in compartments at the cell surface (plasma membrane) and inside the cell (endosome). The trafficking rates correspond to the internalization rates of the different molecular species from the plasma membrane to the endosome and to the recycling rates from the endosome to the plasma membrane. The change between the different molecular species includes their degradation and the formation of the ligand-receptor complexes.

The accuracy of this approximation depends on the characteristic time scales of trafficking with respect to other cellular processes. It has proved to work exceptionally well for the EGFR pathway, which has a fast kinetics reaching maximum activity at ~5 min after stimulation (18). In the case of the TGF-β pathway, the dynamics is substantially slower, with maximum activities at ~60 min, and the agreement with experiments is mostly qualitative. The main reason for these different time scales is that most of the EGF receptors are present in the plasma membrane and they are ready to signal upon the addition of ligand. In the TGF-β pathway, internalization occurs continuously and only about 10% of the receptors are present in the plasma membrane at a given time (17). The remaining 90% of the receptors are internalized in endosomes (11). TGF-β receptors need to be recycled from the endosomes back to the plasma membrane in order to be able to interact with the ligand and this process takes about ~30 min (16).

To take into account the trafficking dynamics, such as that of the TGF-β pathway, in a more detailed way, we describe trafficking by a density $\rho_i(\theta,t)$ for each species along a generalized trafficking coordinate $\theta$ at time $t$ with dynamics given by

$$\frac{\partial \rho_i}{\partial t} = -\frac{\partial J_i}{\partial \theta} + F_i(\rho_1,...,\rho_N),\qquad(2)$$

where $J_i(\theta,t)$ is the flux of the species $i$ and $F_i$ is the functional that describes the change of $\rho_i$ as a result of the reaction between components. The introduction of a trafficking coordinate allows us to consider explicitly the situation in which the properties of the compartments change continuously in a way that cannot naturally be described by compartmentalized chemical reactions.

The flux term is expressed as $J_i = v_i \rho_i - D_i \partial \rho_i / \partial \theta$, which accounts for transport with drift $v_i$ and diffusion $D_i$. Thus, equation (2), without the functional $F_i$, can be interpreted as a Fokker-Planck equation for the probability distribution of the underlying stochastic Brownian motion of the trafficking coordinate of each molecular species (19). The combination of the Fokker-Planck dynamics with the functional $F_i$ implements trafficking in the form of a reaction-diffusion-advection process along the trafficking coordinate.

The functional $F_i$ is the counterpart of the reaction terms in the traditional approach and describes the production, degradation, and interconversion of the molecular species. As we show in detail below, this term has in general an integral form to take into account non-local interactions along the trafficking-coordinate space. Non-local effects arise because two points that are nearby in the cell can be far apart in the trafficking-coordinate space. For instance, receptors close to the cell surface can have very different values of $\theta$ depending on whether they are about to be internalized or they are getting into the plasma membrane. This non-locality is not present in traditional reaction-diffusion and reaction-diffusion-advection approaches for the study of multicellular (20-23) and intracellular (10, 24-26) systems, which take into account spatial rather than trafficking-coordinate inhomogeneities.

We focus explicitly on the coupling between trafficking and signaling in the TGF-β pathway. In this pathway, the information progresses sequentially from the interaction of ligands with transmembrane receptors, through phosphorylation of mediator Smad proteins, to transcriptional responses. The mode of functioning relies on extracellular ligands assembling with type I and type II receptors to form complexes in which type II receptor activates type I receptor by phosphorylating it. Active type I receptors within the complex, in turn, phosphorylate the regulatory Smad proteins in the cytosol, which propagate the signal downstream to transcriptional responses (27, 28).

## Methods

The TGF-β pathway is part of a complex signal transduction network that integrates signals from the 33 known ligands of the TGF-β superfamily (29, 30). There are 7 different type II receptors and 5 different type I receptors that signal through either the Smad2/3 or Smad1/5/8 channels. We analyze explicitly Smad2 phosphorylation through the canonical TGF-β pathway — the ligand TGF-β1 and the receptors Alk5 (type I) and TGFβRII (type II)— in response to changes in ligand concentration. Therefore, we consider three distinct trafficking species: type I receptors, type II receptors, and ligand-receptor complexes. They are indexed by $RI$, $RII$, and $RC$, respectively.

The key elements of the TGF-β pathway we consider (Figure 1) and their implementation in a trafficking-signaling model are expounded below.

## Trafficking coordinate

The trafficking coordinate $\theta$ describes the progression of the molecular species along the trafficking process and its value indicates the cellular compartment location. Explicitly, we consider values of $\theta$ in the interval from 0 to 1. Presence at the plasma membrane (cell surface) corresponds to $\theta \in [0, \theta^M)$, with the molecular species getting into the plasma membrane at $\theta = 0$ and being internalized at $\theta = \theta^M$. The interval $\theta \in [\theta^M, 1)$ is assigned to internalized species and recycling back to the plasma membrane occurs at $\theta = 1$.

## Molecular densities

The distribution of receptors and ligand-receptor complexes in the cell is taken into account by the densities $\rho_i(\theta, t)$, with $i = \{RI, RII, RC\}$, along the trafficking coordinate $\theta$ at time $t$. In terms of these densities, the total numbers of molecules of each species are expressed as $N_i(t) = \int_0^1 \rho_i(\theta, t) d\theta$, and their amounts at the plasma membrane and internalized inside the cell are given by

$$N_i^M(t) = \int_0^{\theta^M} \rho_i(\theta, t) d\theta \tag{3}$$

and

$$N_i^I(t) = \int_{\theta^M}^1 \rho_i(\theta, t) d\theta, \tag{4}$$

respectively.

Ligands induce the formation of receptor complexes with type I and type II receptors. We assume that the formation of complexes is proportional to the ligand concentration, $[l]$, and to the numbers of receptors at the plasma membrane for each of the two receptor types. Explicitly, a receptor of one type with trafficking coordinate $\theta$ at the plasma membrane can interact with all the receptors of the other type that are present at the plasma membrane to form a complex with the ligand. Therefore, the densities of type I and type II receptors decrease at a rate $-k_a[l]\rho_{RI}(\theta,t)I^M(\theta)N^M_{RII}(t)$ and $-k_a[l]\rho_{RII}(\theta,t)I^M(\theta)N^M_{RI}(t)$, respectively. Here, $k_a$ is an effective association constant and $I^M(\theta)$ is an indicator function, defined as

$$I^M(\theta) = \begin{cases} 1 & \text{if } \theta \in [0,\theta^M) \\ 0 & \text{if } \theta \in [\theta^M,1) \end{cases}, \qquad (5)$$

which is 1 when the coordinate $\theta$ is at the plasma membrane and 0 otherwise. This indicator function is used to implement that the reaction takes place only at the plasma membrane. The resulting formation of ligand-receptor complexes is assumed to be distributed uniformly at the plasma membrane, which implies that the density of the complexes increases at a rate $k_a[l]N^M_{RI}(t)N^M_{RII}(t)I^M(\theta)/\int_0^1 I^M d\theta$. The factor $I^M(\theta)/\int_0^1 I^M(\theta)d\theta$, which is $1/\theta^M$ for values of $\theta$ at the plasma membrane and 0 otherwise, transforms the total production rate into the density production rate along the trafficking coordinate.

## Molecular transport

Receptors and ligand-receptor complexes are continuously internalized and recycled back to the plasma membrane (16, 17). In terms of the densities, molecular transport is described by the flux

$$J_i(\theta,t) = v_i \rho_i(\theta,t) - D_i \frac{\partial}{\partial \theta}\rho_i(\theta,t) , \qquad (6)$$

where $v_i$ is the drift and $D_i$ is the diffusion coefficient. Recycling and receptor production are taken into account by the boundary conditions.

For ligand-receptor complexes, which dissociate into their constituent elements before coming back to the plasma membrane, we implement boundary conditions with no complexes at the end of the trafficking coordinate, $\rho_{RC}(1,t) = 0$, and zero flux of complexes at origin, $J_{RC}(0,t) = 0$.

For both types of receptors, complete recycling leads to densities that are the same at both extremes of the trafficking coordinate domain, $\rho_{RI}(0,t) = \rho_{RI}(1,t)$ and $\rho_{RII}(0,t) = \rho_{RII}(1,t)$, which means that the receptors continuously transition between $\theta = 0$ and $\theta = 1$.

The relation between the fluxes at both extremes of the coordinate domain includes recycling from the receptor and also partial recycling from the complexes and receptor production. Taking into account these contributions, we obtain $J_{RI}(0,t) = J_{RI}(1,t) + \alpha J_{RC}(1,t) + r_{RI}$ and $J_{RII}(0,t) = J_{RII}(1,t) + \alpha J_{RC}(1,t) + r_{RII}$ for type I and II receptors, respectively. Here production of receptors is described as a contribution to the fluxes at the origin, where $r_i$ is the receptor production rate. The partial recycling contribution from the complexes, $\alpha J_{RC}(1,t)$, arises because upon dissociation of the ligand-receptor complex, only a fraction $\alpha$ of its constituent receptors become available for signaling at the plasma membrane.

The fact that only a fraction of receptors become available constitutes a form of ligand-induced degradation, which affects only receptors that have been complexed with ligands (16, 17). In addition, receptor degradation has a constitutive contribution, which is the same for free receptors and ligand-receptor complexes (16, 17). It is implemented as $-\gamma \rho_i(\theta,t)$, where $\gamma$ is the degradation rate constant.

## Signaling activity

The connection of the trafficking dynamics with signaling activity is done by considering that Smad phosphorylation is proportional to the number of ligand-receptor complexes that are present in the early endosomes: $k_{phos}\rho_{RC}(\theta^{EE},t)$ (31). Here, $k_{phos}$ is the phosphorylation rate constant of the Smad proteins and $\theta^{EE}$ is the value of the trafficking coordinate at the early endosome.

## Mathematical implementation

We assemble these elements into a mathematical model that describes the dynamics of both the transformation of the different molecular species into each other and their trafficking in the cell. The resulting equations are

$$\frac{\partial}{\partial t}\rho_{RI} = -\frac{\partial}{\partial \theta}J_{RI} - \gamma\rho_{RI} - k_a[l]\rho_{RI}I^M N_{RII}^M,$$

$$\frac{\partial}{\partial t}\rho_{RII} = -\frac{\partial}{\partial \theta}J_{RII} - \gamma\rho_{RII} - k_a[l]\rho_{RII}I^M N_{RI}^M, \quad (7)$$

$$\frac{\partial}{\partial t}\rho_{RC} = -\frac{\partial}{\partial \theta}J_{RC} - \gamma\rho_{RC} + k_a[l]N_{RI}^M N_{RII}^M \frac{I^M}{\int_0^1 I^M d\theta},$$

with the boundary conditions:

$$J_{RI}(0,t) = J_{RI}(1,t) + \alpha J_{RC}(1,t) + r_{RI}, \quad \rho_{RI}(0,t) = \rho_{RI}(1,t),$$
$$J_{RII}(0,t) = J_{RII}(1,t) + \alpha J_{RC}(1,t) + r_{RII}, \quad \rho_{RII}(0,t) = \rho_{RII}(1,t), \quad (8)$$
$$J_{RC}(0,t) = 0, \quad \rho_{RC}(1,t) = 0.$$

The signaling activity is given by

$$\frac{dS_p}{dt} = k_{phos}\rho_{RC}(\theta^{EE},t) - \lambda S_p, \quad (9)$$

which corresponds to the dynamics of phosphorylated Smad proteins, $S_p$, with dephosphorylation rate constant $\lambda$.

## Numerical integration

The resulting system of generalized reaction-diffusion-advection equations for the trafficking-signaling dynamics cannot generally be solved analytically. To obtain the dynamics we integrate the equations numerically with an explicit first-order upwind scheme on a 200-point mesh along the trafficking coordinate with a variable time step (32). The same time steps are used to integrate the ordinary differential equation of the signaling dynamics with a first-order Euler method.

## Compartmental model

It is possible to obtain a compartmental approximation to the trafficking-coordinate description by considering the total numbers of the different molecular species at the plasma membrane and inside the cell. The first step is to integrate the set of equations (7) with respect to the coordinate $\theta$ along both the plasma membrane, $\theta \in [0, \theta^M)$, and the internalized region, $\theta \in [\theta^M, 1)$. For the plasma membrane, using the relationships $\int_0^{\theta^M} (\partial \rho_i / \partial t) d\theta = dN_i^M / dt$ and $\int_0^{\theta^M} (\partial J_i / \partial \theta) d\theta = J_i(\theta^M, t) - J_i(0, t)$ and taking into account the boundary conditions to eliminate the fluxes at $\theta = 0$, we obtain

$$dN_{RI}^M / dt = \alpha J_{RC}(1,t) + r_{RI} + J_{RI}(1,t) - J_{RI}(\theta^M, t) - \gamma N_{RI}^M - k_a[l] N_{RI}^M N_{RII}^M,$$
$$dN_{RII}^M / dt = \alpha J_{RC}(1,t) + r_{RII} + J_{RII}(1,t) - J_{RII}(\theta^M, t) - \gamma N_{RII}^M - k_a[l] N_{RI}^M N_{RII}^M, \quad (10)$$
$$dN_{RC}^M / dt = -J_{RC}(\theta^M, t) - \gamma N_{RC}^M + k_a[l] N_{RI}^M N_{RII}^M.$$

For the internalized region, the relationships $\int_{\theta^M}^1 (\partial \rho_i / \partial t) d\theta = dN_i^I / dt$ and $\int_{\theta^M}^1 (\partial J_i / \partial \theta) d\theta = J_i(1,t) - J_i(\theta^M, t)$ lead to

$$dN_{RI}^I / dt = J_{RI}(\theta^M, t) - J_{RI}(1,t) - \gamma N_{RI}^I,$$
$$dN_{RII}^I / dt = J_{RII}(\theta^M, t) - J_{RII}(1,t) - \gamma N_{RII}^I, \quad (11)$$
$$dN_{RC}^I / dt = J_{RC}(\theta^M, t) - J_{RC}(1,t) - \gamma N_{RC}^I.$$

These equations indicate that there is no exact connection of the trafficking-coordinate model with the two-compartment model. The two-compartment model implicitly considers that all the molecular species have the same properties inside any of the two compartments, which corresponds to replacing the fluxes by the drift multiplied by the average densities. Therefore, the heuristic assumptions of the two-compartment model $J_i(1,t) \rightarrow v_i N_i^I / (1-\theta^M)$ and $J_i(\theta^M, t) \rightarrow v_i N_i^M / \theta^M$ lead to a closed set of equations with recycling $e_i = v_i / (1-\theta^M)$ and internalization $a_i = v_i / \theta^M$ rate constants for the molecular species at the plasma membrane,

$$\frac{dN_{RI}^M}{dt} = r_{RI} - k_a[l] N_{RI}^M N_{RII}^M + \alpha e_{RC} N_{RC}^I + e_{RI} N_{RI}^I - (a_{RI} + \gamma) N_{RI}^M,$$
$$\frac{dN_{RII}^M}{dt} = r_{RII} - k_a[l] N_{RI}^M N_{RII}^M + \alpha e_{RC} N_{RC}^I + e_{RII} N_{RII}^I - (a_{RII} + \gamma) N_{RII}^M, \quad (12)$$
$$\frac{dN_{RC}^M}{dt} = k_a[l] N_{RI}^M N_{RII}^M - (a_{RC} + \gamma) N_{RC}^M,$$

and inside the cell,

$$\frac{dN^I_{RI}}{dt} = a_{RI} N^M_{RI} - (e_{RI} + \gamma) N^I_{RI},$$

$$\frac{dN^I_{RII}}{dt} = a_{RII} N^M_{RII} - (e_{RII} + \gamma) N^I_{RII}, \qquad (13)$$

$$\frac{dN^I_{RC}}{dt} = a_{RC} N^M_{RC} - (e_{RC} + \gamma) N^I_{RC}.$$

The signaling activity is obtained with the assumption $\rho_{RC}(\theta^{EE}, t) \to N^I_{RC}/(1-\theta^M)$, which leads to

$$\frac{dS_p}{dt} = \tilde{k}_{phos} N^I_{RC} - \lambda S_p, \qquad (14)$$

where the effective phosphorylation rate constant of the Smad proteins is given by $\tilde{k}_{phos} = k_{phos}/(1-\theta^M)$.

## Parameter values

We express the trafficking coordinate, the density of the receptors, and phosphorylated Smad in terms of dimensionless quantities by normalizing them by the length of a trafficking round, the number of receptors produced during one hour, and the maximum signaling activity, respectively. As time units, we use hours. To obtain the values of the different parameters for the system of equations, we started with the experimentally available values.

The total trafficking time of the receptors, from internalization to recycling back to the plasma membrane, has been measured to be about 30 min (16). We selected the drifts $v_{RII} = v_{RI} = v_{RC} = 2 \, \text{h}^{-1}$ to capture this characteristic time in which receptors undergo on average two rounds of internalization and recycling in an hour. Typical receptor internalization times are 3 min (17), which in the model corresponds to $\theta^M = 0.1$ so that it matches the characteristic transit time of the receptors at the plasma membrane, $\theta^M / v_i = 0.05 \, \text{h} = 3 \, \text{min}$. Similarly, we selected $\theta^{EE} = 0.35$ to implement that signaling from the early endosomes happens at around 10 min (31). The values of diffusion coefficients have not been measured experimentally. We have selected $D_{RII} = D_{RI} = D_{RC} = 0.05 \, \text{h}^{-1}$ so that the effects of diffusion in the internalization process are slightly smaller than those of advection. With these parameter values, the corresponding Péclet number is $\theta^M v_i / D_i = 4$.

By definition of the dimensionless quantities, the number of receptors produced during an hour is one and therefore we have $r_{RII} = r_{RI} = 1 \, \text{h}^{-1}$. To mimic the experimental conditions that change from zero to saturating values of the ligand concentration, we have chosen $k_a[l] = 10^4 \Theta(t) \, \text{h}^{-1}$, where $\Theta(t)$ is the Heaviside unit step function, which is sufficiently high, for $t > 0$, to induce the maximal formation of ligand-receptor complexes at the plasma membrane.

For the dephosphorylation rate of Smad2, we select $\lambda = 12.0 \, \text{h}^{-1}$ to account for the experimental observations that upon inhibiting the kinase activity of receptor type I phosphorylated Smad2 gets dephosphorylated with a characteristic time of 5 min (33).

The constitutive degradation rate of the receptors and fraction of recycled ligand-receptor complexes are obtained by adjusting their value to reproduce the signaling dynamics of Smad2. The values obtained for HaCat cell lines (33) are $\gamma = 0.0714 \, \text{h}^{-1}$ and $\alpha = 0.896$. The corresponding value for the phosphorylation rate constant of Smad2 is $k_{phos} = 0.83 \, \text{h}^{-1}$.

Based on this set of parameter values, different experiments can be reproduced by adjusting only the three independent parameters $v_{RII} = v_{RI}$, $v_{RC}$, and $\alpha$, which determine the global trafficking properties, to reduce the square error between the model and experimental results.

## Results

We focus on the prototypical experimental conditions that consider the response to sustained changes in TGF-β concentration that suddenly increase from zero to saturating values and are kept constant afterward. The general behavior of this type of response shows partial adaptation after reaching a maximum of activity.

The trafficking-coordinate model described by equations (5)-(9) implements the continuous dynamics followed by a population of receptors along the trafficking process (Figure 2). Explicitly, before stimulation of the pathway (negative times in Figure 2), when the system has reached the steady state with zero ligand concentration, type I and II receptors are distributed almost uniformly along the trafficking coordinate and there are no complexes. Upon addition of saturating concentrations of ligand, type I and II receptors at the cell surface assemble into ligand-receptor complexes, leading to the depletion of both existing type I and II receptors and newly generated receptors from production and recycling of the complexes. Internalized complexes, traffic inside the cell, phosphorylate the Smad proteins at the early endosomes ($\theta = \theta^{EE}$), and their constituent components keep trafficking until they reach $\theta = 1$, where a fraction of them are recycled back into type I and II receptors.

Comparison with the experimental data shows that this approach, which takes into account the detailed trafficking dynamics (Figure 2), can accurately reproduce the observed time courses of phosphorylated Smad upon addition of saturating ligand concentrations (Figure 3A). The time courses include a lag phase of about 10 min duration, a sharp rise to maximum peak activity up to 30 min after ligand addition, and a sharp transition to slowly decreasing signaling activity. To obtain the values of the different parameters for the system of equations, we started with the general set of parameter values inferred from the experimentally available data, and then we adjusted the drifts $v_{RII} = v_{RI}$ and $v_{RC}$ to reduce the square error between the model and experimental results.

To validate the model and verify its ability to predict the effects of perturbations in the system without free parameters, we considered the effects of the drug cycloheximide. This drug is an inhibitor of protein biosynthesis in eukaryotic organisms.

Its most direct effect in the pathway is to stop production of receptors. When receptor production is set to zero just before the ligand is added, the model is able to capture the main features of the experimental data for cycloheximide treatment (Figure 3B). The agreement with the experimental data, in which only small differences are observed, is highly remarkable because it is an *ab initio* prediction of the model without any fitting to experimental data for cycloheximide treatment.

We investigated the origin of the small differences observed between the model and the experimental data. Since protein production uses the trafficking apparatus of the cell, we adjusted the drifts of the trafficking fluxes to improve even further the agreement. We observed that the small differences (Figure 3B) virtually disappear for the new set of drifts (Figure 3C), which are faster than without cycloheximide, especially for ligand-receptor complexes. The observed increases in velocity are consistent with a shift of the trafficking machinery of the cell to signaling from shuttling to their cellular location of newly produced proteins.

The spatiotemporal dynamics of the receptors and complexes during the early stages of signaling exhibits a behavior that is not captured with standard two-compartment reaction equations. The main feature is the trafficking of complexes that reach the early endosomes $\theta^{EE} = 0.35$ after ~10 min of ligand addition (Figure 2). The consequence is a sharp rise of the signaling activity (see also detail in Figures 3D and 3E), which cannot be achieved by the traditional two-compartment model.

Indeed, the dynamics of the two-compartment approximation given by equations (12)–(14) describes in detail only the long term behavior of the systems without (Figure 4A) and with (Figures 4B and 4C) cycloheximide treatment but is unable to capture the abrupt rise of the signaling activity observed for short times (see also detail in Figures 4D and 4E). The slow rise in signaling activity of the two-compartment model leads its signaling activity to peak at about 60 min instead of the observed 30 min.

In the system with cycloheximide (Figure 3E), in addition, there is a rippling effect on the top of the response. Such superposed oscillation on the top of the main raise and decay of the response is present in the experimental data of other systems, like the EGF receptor pathway (Figure 2E of Ref. (34)) but it has never been interpreted before. In our analysis, it originates from the differences between the drifts of receptors and ligand-receptor complexes in the trafficking-coordinate model but it is not present in the two-compartment model (Figure 4E) or in the trafficking-coordinate model when the drift differences are much smaller (Figures 3A, 3B, and 3C).

A key property of trafficking is its ability to change the qualitative behavior of the system. There is now ample evidence that there are regulated mechanisms, as for instance glycosylation of EGF and TGF-β receptors, that change the trafficking properties to modify both the signaling dynamics and the physiological outcomes (35). We focus on cell lines in which the duration of the signaling activity correlates with the physiological outcomes triggered by TGF-β. The ligand TGF-β is especially important in cancer because of its role as a cell growth suppressor. In particular, epithelial cells that are sensitive to the antiproliferative effects of TGF-β (HaCaT and BxPC3 cell lines) have sustained activity of more than 4-6 h (36). In contrast, pancreatic cancer cell lines (PT45 and Panc-1), which are resistant to TGF-β-induced growth arrest, show short transient activity of about 1-2 h (36).

The results of our analysis (Figure 5) indicate that these differences in signaling between non-cancerous and cancerous cell lines are captured in detail by changes in the trafficking patterns of the receptors. The responses to a permanent change in ligand concentration are characterized on one extreme type of behavior by a transient increase in signaling activity that returns to pre-stimulus levels for high ligand-induced degradation and on the other extreme by a permanently elevated level of signaling activity for low ligand-induced degradation. At these two extremes, there are the cell lines PT45 (Figure 5A) and BxPC3 (Figure 5D) with complete degradation and with full recycling, respectively, of the ligand-receptor complexes. Interestingly, the inferred trafficking velocities are slower for cancerous cell lines, which is consistent with a shift of the trafficking machinery of the cell to shuttling newly produced proteins needed for growth, in agreement with the opposite trend we observed for treatment with cycloheximide (Figures 3C and 3E).

## Discussion

The TGF-β pathway integrates signals from the 33 known ligands of the TGF-β superfamily (29, 30). It encompasses many processes that affect the propagation of the signal, such as nucleo-cytoplasmic shuttling of Smad proteins, Smad dephosphorylation, and ligand availability. Several recent studies based on chemical kinetics have incorporated many of these regulatory levels into different quantitative models of the TGF-β pathway that provide insights into many aspects of the pathway behavior (37-41).

Here, we have focused on the most widely studied member of TGF-β superfamily, for ligand concentrations that change from zero to saturating values, in systems with fast phosphorylation-dephosphorylation and nuclear import-export time scales. Therefore, the system behavior is dominated by trafficking itself and we do not need to take into account these extra levels explicitly.

The focus on trafficking-dominated conditions allowed us to uncover that the control of signaling by intracellular trafficking follows a well-defined kinetics along a trafficking coordinate. Explicitly, we have shown that the combination of Fokker-Planck and chemical reaction kinetics into a generalized reaction-diffusion-advection model, in contrast to previous two-compartment models, accurately captures the signaling dynamics altogether for short, intermediate, and long time scales. These results cover the range from purely transient to permanent responses, including those of the transition from cancerous to non-cancerous cell types, and can predict, without free parameters, the effects of drugs in a way that closely matches the observed behavior.

The newly introduced concept of trafficking coordinate makes it possible to accurately recapitulate the distinct signaling dynamics of different cell types by taking into account the detailed physical properties of the trafficking processes. Our results thus offer a solid starting point to couple detailed physical transport and chemical reaction kinetics into predictive quantitative frameworks that faithfully integrate multiple control layers of signal transduction and processing.

## Acknowledgments

This work was supported by the MICINN under grant FIS2009-10352 (J.M.G.V.) and the University of California, Davis (L.S.).

## Figure legends

**Figure 1:** Two-compartment and trafficking-coordinate descriptions of the TGF-β signal transduction network. The components of the system are type I (TβRI) and II (TβRII) receptors, ligand-receptor complexes (TβRC), ligands, and Smad proteins. In the traditional *two-compartment description* (left), receptors can be present at the plasma membrane (cell surface) or in endosomes (internalized). Receptors and ligand-receptor complexes traffic between these two compartments. Only internalized ligand-receptor complexes phosphorylate Smad proteins. The ligand forms complexes with type I and type II receptors at the plasma membrane. Receptors can undergo constitutive degradation, independently of whether they are ligand-bound and can recycle back to the plasma membrane. New receptors are constantly produced as a result of gene expression. The *trafficking-coordinate description* (right) considers the progression of the different receptor species during the trafficking process. Presence at the plasma membrane corresponds to $\theta \in [\theta^0, \theta^M)$, with the molecular species getting into the cell surface at $\theta = \theta^0 = 0$ and being internalized at $\theta = \theta^M$. Internalized species belong to the interval $\theta \in [\theta^M, \theta^1)$ and a fraction of them are recycled back to the plasma membrane at $\theta = \theta^1 = 1$. Smads are phosphorylated at $\theta = \theta^{EE}$ in the early endosomes. In both cases, the phosphorylated Smads are the signal, which can then translocate into the nucleus where they act as transcriptional regulators of about 300 target genes.

**Figure 2:** Spatiotemporal dynamics of trafficking upon TGF-β stimulation at time 0. The temporal evolution (vertical axis) from time -0.25 to 1.25 hours of the densities of (*A*) type I receptor, $\rho_{RI}$, and (*B*) ligand-receptor complex, $\rho_{RC}$, along the trafficking coordinate (horizontal axis) shows that the distributions of the different molecular species along the trafficking-coordinate space are highly inhomogeneous during the early stages of TGF-β signaling. The cell surface corresponds to the interval $\theta \in [0, 0.1)$ and Smad phosphorylation occurs in the early endosomes at $\theta = 0.35$. The dynamics is given by Eqs. (5)-(9). The values of the parameters are $\theta^M = 0.1$, $\theta^{EE} = 0.35$, $v_{RII} = v_{RI} = 1.96 \text{ h}^{-1}$, $v_{RC} = 1.90 \text{ h}^{-1}$, $D_{RII} = D_{RI} = D_{RC} = 0.05 \text{ h}^{-1}$, $k_a[l] = 10^4 \Theta(t) \text{ h}^{-1}$, $\gamma = 0.0714 \text{ h}^{-1}$, $r_{RII} = r_{RI} = 1 \text{ h}^{-1}$, $\alpha = 0.896$, $k_{phos} = 0.83 \text{ h}^{-1}$, and $\lambda = 12.0 \text{ h}^{-1}$.

**Figure 3:** Dynamics of the signaling activity upon TGF-β stimulation. The activity from the trafficking-coordinate model (solid lines) corresponds to $S_p$ from Eqs. (5)-(9). The experimental data (symbols) has been obtained from phosphorylated Smad2 western blots of Figure 1A of Ref. (33) after quantification and normalization with the software ImageJ (http://rsbweb.nih.gov/ij/). (*A*) Activity for HaCaT cells upon increasing at time zero the TGF-β ligand concentration from zero to saturating values. The time units are hours and the trafficking coordinate, the density of the receptors, and phosphorylated Smad have been made dimensionless by normalizing them by the length of a trafficking round, the number of receptors produced during an hour, and the maximum signaling activity, respectively. The values of the parameters are the same as in Figure 2. (*B*) Activity for HaCaT cells upon stimulation with TGF-β after pre-treatment with cycloheximide. The values of the parameters are the same as in Figure 2 but receptor production is switched off at time zero: $r_{RII} = r_{RI} = 1 - \Theta(t)$. (*C*) Activity for HaCaT cells upon stimulation with TGF-β after pre-treatment with cycloheximide for inferred trafficking and signaling rates $v_{RII} = v_{RI} = 2.34 \text{ h}^{-1}$, $v_C = 2.96 \text{ h}^{-1}$, and $k_{phos} = 0.76 \text{ h}^{-1}$. The values of the other parameters are the same as in Figure 2. (*D*) Detail of the signaling activity of panel A. (*E*) Detail of the signaling activity of panel C.

**Figure 4:** Dynamics of the two-compartment model approximation. The signaling activity (solid lines) of the two-compartment model is given by $S_p$ from Eqs. (12)-(14). The experimental conditions and data (symbols) are the same as in Figure 3. The values of the parameters were obtained from those of Figure 2 using $e_i = v_i / (1 - \theta^M)$ for recycling and $a_i = v_i / \theta^M$ for internalization rate constants: $a_{RII} = a_{RI} = 19.6 \text{ h}^{-1}$, $a_{RC} = 19.0 \text{ h}^{-1}$, $e_{RII} = e_{RI} = 2.18 \text{ h}^{-1}$, $e_{RC} = 2.11 \text{ h}^{-1}$, $k_a[l] = 10^4 \Theta(t) \text{ h}^{-1}$, $\gamma = 0.0714 \text{ h}^{-1}$, $r_{RII} = r_{RI} = 1 \text{ h}^{-1}$, $\alpha = 0.896$, $\tilde{k}_{phos} = k_{phos} / (1 - \theta^M) = 0.92 \text{ h}^{-1}$, and $\lambda = 12.0 \text{ h}^{-1}$. (*A*) Activity for HaCaT cells upon increasing at time zero the TGF-β ligand concentration from zero to saturating values. (*B*) Activity for HaCaT cells upon stimulation with TGF-β after pre-treatment with cycloheximide. The values of the parameters are the same as in Figure 2 but receptor production is switched off at time zero: $r_{RII} = r_{RI} = 1 - \Theta(t)$. (*C*) Activity for HaCaT cells upon stimulation with TGF-β after pre-treatment with cycloheximide for inferred trafficking and signaling rates $a_{RII} = a_{RI} = 2.34 \text{ h}^{-1}$, $a_C = 2.96 \text{ h}^{-1}$, $e_{RII} = e_{RI} = 2.60 \text{ h}^{-1}$, $e_{RC} = 3.28 \text{ h}^{-1}$, and $k_{phos} = 0.84 \text{ h}^{-1}$. The values of the other parameters are the same as in Figure 2. (*D*) Detail of the signaling activity of panel A. (*E*) Detail of the signaling activity of panel C. The dynamics is computed numerically with an implicit backward differentiation solver with orders 1 through 5.

**Figure 5:** Control of the signaling activity for cancerous and non-cancerous cell lines. The general response to sustained changes in TGF-β concentration shows partial adaptation after reaching a maximum of activity. Different trafficking properties select the balance between transient and permanent contributions. In all panels, the TGF-β concentration is increased at time zero from zero to saturating values and kept constant afterwards, as in Figure 3. The only differences among different panels are the trafficking rates and the extent of ligand-induced degradation. The activity from the model (solid lines) corresponds to $S_p$ from Eqs. (5)-(9). The experimental data (symbols) has been obtained from phosphorylated Smad2 western blots of Figure 4a of Ref. (36) after quantification and normalization with the software ImageJ (http://rsbweb.nih.gov/ij/). (*A*) Activity of PT45 cells with $v_{RII} = v_{RI} = 0.76$, $v_{RC} = 0.70$, and $\alpha = 0.0$. (*B*) Activity of Panc-1 cells with $v_{RII} = v_{RI} = 0.76$, $v_{RC} = 0.80$, and $\alpha = 0.5$. (*C*) Activity of HaCaT cells with $v_{RII} = v_{RI} = 1.96 \text{ h}^{-1}$, $v_{RC} = 1.90 \text{ h}^{-1}$, and $\alpha = 0.896$. (*D*) Activity of BxPC3 cells with $v_{RII} = v_{RI} = 1.96 \text{ h}^{-1}$, $v_{RC} = 1.90 \text{ h}^{-1}$, and $\alpha = 1.0$. The values of the other parameters are as in Figure 3A. Note that the values of the parameters for HaCaT cells in panel C and Figure 3A are the same.

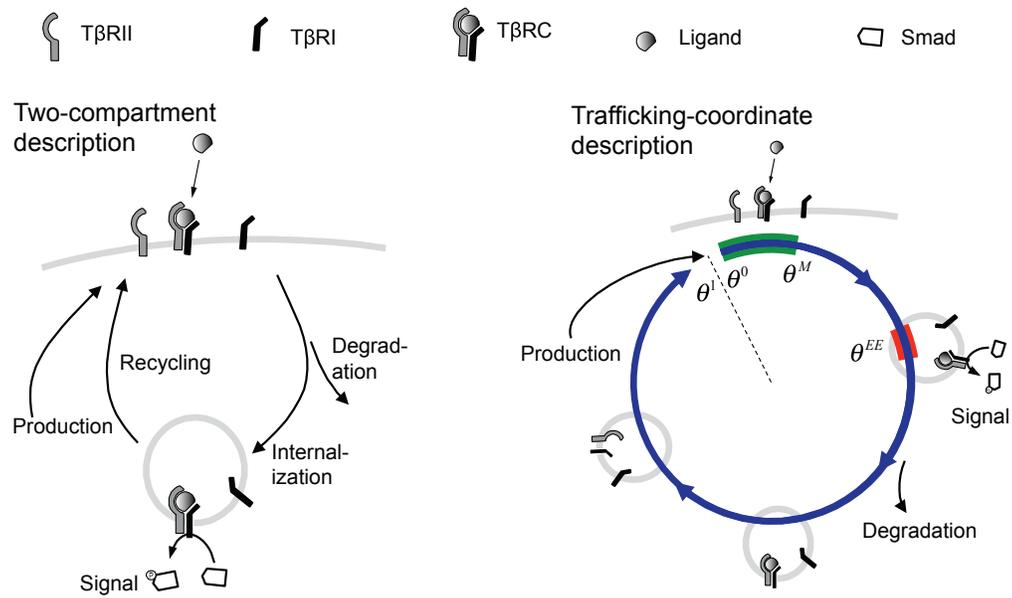

Figure 1

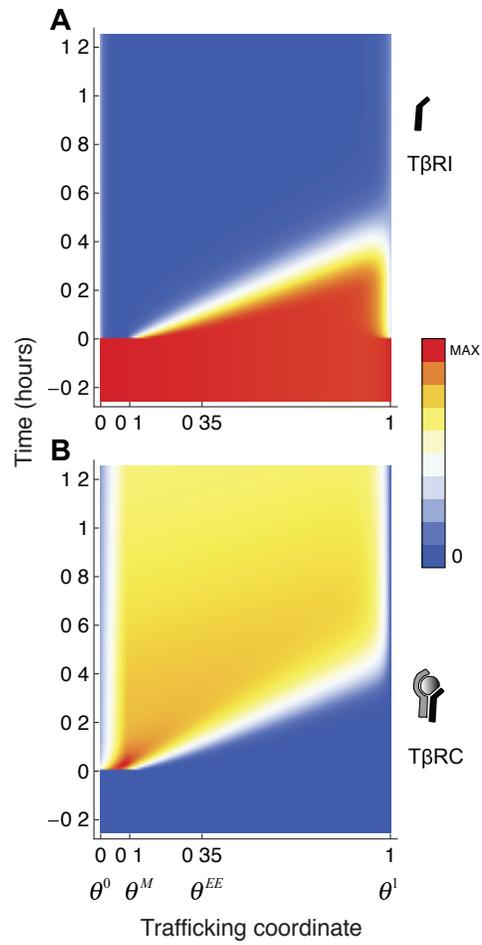

Figure 2

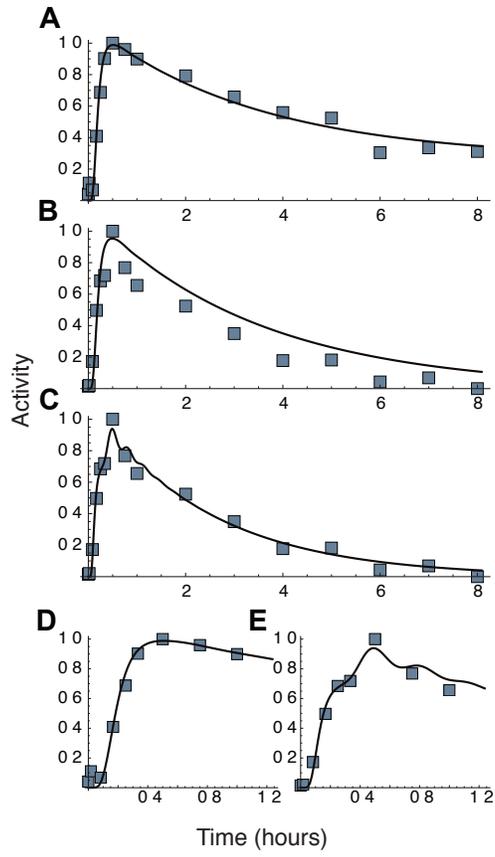

Figure 3

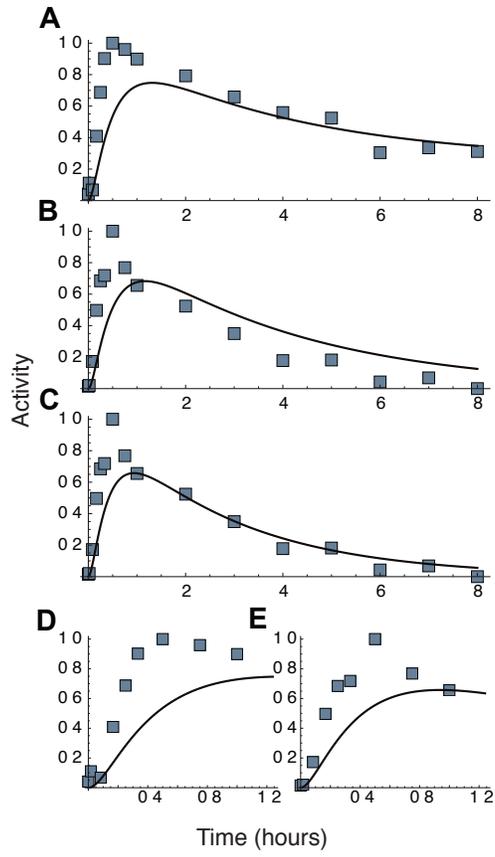

Figure 4

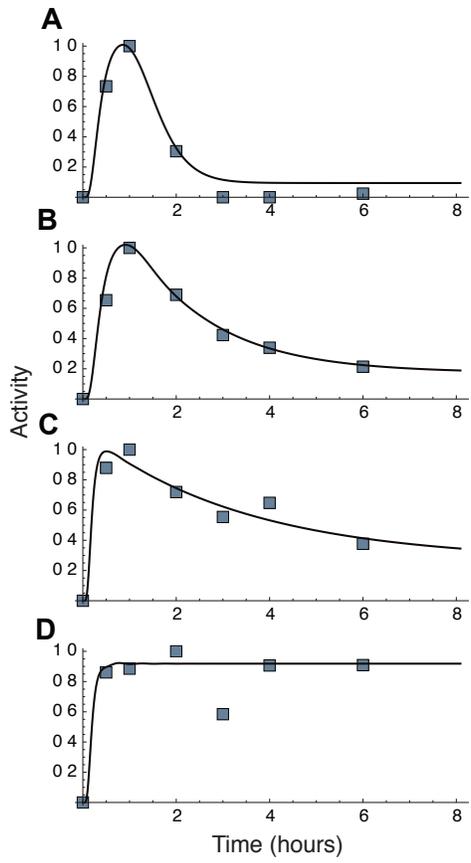

Figure 5